\documentclass[aps,pre,twocolumn,epsfig,floats,superscriptaddress]{revtex4}
\usepackage{amsmath}
\usepackage{color}
\usepackage{graphicx}
\usepackage{amsfonts}
\usepackage{mathrsfs}
\usepackage{amsmath}
\usepackage{amssymb}
\usepackage[utf8x]{inputenc}
\usepackage{times}
\usepackage{hyperref}
\hypersetup{
  colorlinks=true,
  citecolor=blue,
  linkcolor=blue,
  urlcolor=blue}

\begin{document}

\title{Fingerprint of chaos and quantum scars in kicked Dicke model: An out-of-time-order correlator study}
 
\author{Sudip Sinha}
\affiliation{Indian Institute of Science Education and Research Kolkata, Mohanpur, Nadia 741246, India}

\author{Sayak Ray}
\affiliation{Physikalisches Institut, Universit\"at Bonn, Nussallee 12, 53115 Bonn, Germany}

\author{Subhasis Sinha}
\affiliation{Indian Institute of Science Education and Research Kolkata, Mohanpur, Nadia 741246, India}
 
\date{\today}

\begin{abstract}

We investigate the onset of chaos in a periodically kicked Dicke model (KDM), using the out-of-time-order correlator (OTOC) as a diagnostic tool, in both the oscillator and the spin subspaces. In the large spin limit, the classical Hamiltonian map is constructed, which allows us to investigate the corresponding phase space dynamics and to compute the Lyapunov exponent. We show that the growth rate of the OTOC for the canonically conjugate coordinates of the oscillator is able to capture the Lyapunov exponent in the chaotic regime. The onset of chaos is further investigated using the saturation value of the OTOC, that can serve as an alternate indicator of chaos in a generic interacting quantum system. This is also supported by a system independent effective random matrix model. We further identify the quantum scars in KDM and detect their dynamical signature by using the OTOC dynamics. The relevance of the present study in the context of ongoing cold atom experiments is also discussed.

\end{abstract}

\maketitle

\section{Introduction}

In the recent years, diagnosis of chaos in quantum many body systems has received much attention in the context of ergodicity related to thermalization of isolated systems \cite{Polkovnikov16, Santos16}. Unlike the classical systems, where chaos leads to mixing in phase space \cite{Casati, Ruelle85}, {\it eigenstate thermalization hypothesis} (ETH) has been put forward to understand the thermalization in closed quantum systems \cite{Deutsch91, Srednicki94} and its connection with random matrix theory (RMT) has been explored extensively \cite{Izrailev90, Rigol10, Rigol16}. In this context, driven quantum systems have drawn a lot of interest due to its intriguing ergodic behavior which can lead to infinite temperature thermalization \cite{Rigol14, Fazio15, Jessen09, Neill16}. In contrast, there are other interacting systems which do not thermalize and remain localized---a phenomena known as many body localization (MBL) \cite{Huse13, Huse15, Altman18, Bloch15}. These localized systems can also delocalize in the presence of a periodic drive, which has been experimentally observed \cite{Bloch17}, as well theoretically investigated in an attempt to understand the role of underlying chaotic dynamics behind such phenomena \cite{Ray18, RayOTOC18}.

In addition to MBL, there exist other factors which can hinder ergodicity too; one of them being the non-ergodic multifractal quantum states giving rise to anomalous thermalization \cite{Lev16, Altland19, Ray19}. Recently, it has been found that there are certain states of an interacting quantum system, which exhibit {\it athermal} behavior and revival phenomena. These states are attributed to {\it many body quantum scar} (MBQS), which has been observed in recent experiments on a chain of Rydberg atoms \cite{Lukin17} and dipolar quantum gas \cite{Kao20}, as well their appearance in many body systems have been analyzed theoretically in a series of recent works \cite{Abanin18, Sinha20, Mondal20, Sengupta20}. Originally, quantum scar has been identified as a reminiscence of an unstable periodic orbit in a non-interacting chaotic system \cite{Heller84}. However, due to the absence of a phase space trajectory in a generic many particle system, a correspondence of MBQS with a classically unstable periodic orbit as well as a connection between thermalization of a quantum many body system and the classical phase space mixing is subtle to date. 

Chaos in classical systems is a fascinating dynamical phenomena giving rise to butterfly effect---an exponential sensitivity towards initial conditions, which is quantified by Lyapunov exponent (LE) \cite{Casati, Ruelle85}. On the contrary, in quantum systems, there is no notion of a phase space, and thus such quantification is not straightforward. However, according to Bohigas-Giannoni-Schmit (BGS) conjecture, the signature of chaos in quantum system can be found from the spectral statistics \cite{BGS}.
Such a fingerprint of chaos has been investigated extensively in quantum systems \cite{Haake} as well as in the presence of periodic kicking e.g. in kicked rotor, kicked top, coupled kicked top models, which also have suitable classical limit that is useful for analyzing their underlying phase space behavior \cite{Izrailev90, Robnik13, Haake87, Deutsch08, Backer20}. The Dicke model (DM) and a time-dependent variant of DM with periodic kicking are another examples of the same class, where the role of underlying chaos observed both in the classical phase space and in spectral statistics have been explored to understand thermalization of these simple interacting models \cite{Altland12, Ray16}. Although, the global chaotic behavior in a quantum system is reflected in its spectral statistics, there is no direct way to quantify its degree, as well to probe the local phase space behavior of a generic many body system, where semiclassical limit is absent.

Interestingly, recent studies show that the out-of-time-order correlator (OTOC) can serve as a tool to detect information scrambling and chaos in quantum systems, revealing quantum butterfly effect \cite{Larkin68, Swingle16, Patel17, Galitski17, Hayden16, Knap17, Dhar18, Lev17}.  
It was first introduced in the context of disordered superconductor \cite{Larkin68}, and later on, it became popular for the Sachdev-Ye-Kitaev (SYK) model \cite{Sachdev93, Kitaev14} which is believed to have a connection with black hole thermalization giving rise to an upper bound of Lyapunov exponent \cite{Shenker14, Maldacena16}. Till date, a number of theoretical studies have used different dynamical and spectral properties of OTOC in the short as well as in long time scales, as an indicator of chaos and ergodicity of many body systems \cite{Galitski19, Garcia18, Garcia20, Santos20, Arul20, Demler16, Fradkin17, Alet18}. Such a quantity has also been measured experimentally in trapped ions and spin systems \cite{Rey17, Du17}. However, a direct correspondence with the Lyapunov exponent relies on an appropriate semiclassical limit, which is normally absent in a generic interacting quantum system. In this paper, we propose an alternate measure, the long-time saturation value of OTOC, which can serve as an indicator for the onset of chaos. Such a measure has been used previously to capture the many body localized (MBL) to thermal phase transition in the presence of periodic drive \cite{RayOTOC18}. 

The purpose of the present work is to analyze the onset of chaos from the dynamics of OTOC, both from its growth and long-time saturation as a case study in `kicked Dicke model' (KDM). The model consists of a large spin which interacts periodically with a bosonic (oscillator) mode and has a suitable classical limit, which allows us to study the phase space dynamics as well to extract Lyapunov exponent from the growth of out-of-time-order correlation between canonically conjugate position and momenta of the oscillator. The degree of chaos is alternatively captured from the saturation value of the OTOC which is particularly studied in the spin sector using the spin operators. Moreover, we identify the dynamical signature of `quantum scars' present in the KDM from the proposed measures of OTOC.

The paper is organized as follows. First, we introduce the KDM and describe the OTOC as a diagnostic tool of chaos in Sec.\ \ref{model}. This is followed by a review on the classical dynamics and spectral statistics of the KDM in Sec.\ \ref{recap}. Next, we discuss the OTOC dynamics and explore its connection with chaos both in the bosonic and spin sectors in Sec.\ \ref{OTOC}. Also, the universal behavior of the OTOC dynamics and the crossover to chaos is examined by constructing an effective random matrix model. Using such OTOC toolbox, dynamical signature of the possible quantum scar states of the KDM are identified in Sec.\ \ref{scar}. Finally, we summarize our work and discuss the relevant experiments in Sec.\ \ref{conclu}. 

\section{Kicked Dicke Model and the OTOC toolbox}
\label{model}

Originally, the Dicke model was introduced to describe a system of $N$ two level atoms represented by spins of angular momentum $1/2$ with energy gap $\hbar \omega_0$, interacting with a single cavity mode of energy $\hbar \Omega$. It is a prototype model which exhibits quantum phase transition from normal to superradiant phase \cite{Dicke, Brandes03}. 
Dicke model has been realized in cavity QED experiment by coupling a Bose-Einstein condensate (BEC) with a cavity mode \cite{Esslinger10, Esslinger13, Hemmerich15}, or using cavity assisted Raman transitions \cite{Barrett14}. Various non-equilibrium dynamics of Dicke model both at the quantum level or close to the classical limit has been explored over the recent past \cite{Keeling19}.  

At zero temperature, such a system can be described collectively by a large spin $S=N/2$, which follows from $\vec{S}_i=\sum_{j=1}^N \hat{\sigma}_i^j/2$ ($\hat{\sigma}_{x,y,z}$ are the Pauli spin matrices), interacting with a bosonic mode of energy $\hbar \Omega$ \cite{Brandes03}. In the presence of periodic kicking in the spin-cavity interaction, the time-dependent Hamiltonian can be written as ($\hbar=1$),
\begin{subequations}
\begin{eqnarray}
\hat{H}(t) &=& \hat{H}_0 + \hat{H}_c(t), \label{Dicke_t} \\
\hat{H}_0 &=& \omega _0 \hat{S}_z + \Omega \hat{a}^{\dagger}\hat{a} \label{H_0} \\
\hat{H}_c(t) &=& \frac{\lambda _0}{\sqrt{S}} (\hat{a}^{\dagger} + \hat{a})\hat{S}_x \sum _{n=-\infty}^{\infty}\delta (t-nT), \label{Dicke_Hct}
\end{eqnarray}
\label{Dicke_hamt}
\end{subequations}
where, the kicking is characterized by its strength $\lambda_0$ and time period $T$, $\hat{a} (\hat{a}^{\dagger})$ represent the bosonic annihilation (creation) operators and $n$ denotes the number of kicking. For numerical computations, we truncate the bosonic number state at $N_{\rm max}$ giving rise to the total Hilbert space dimension $\mathcal{N}=(2S+1)(N_{\rm max}+1)$. The bosonic mode can be alternatively characterized by canonically conjugate position and momenta of a quantum harmonic oscillator, $\hat{Q}_H = \hat{Q}/\sqrt{m\Omega}$ and $\hat{P}_H = \sqrt{m\Omega}\hat{P}$ respectively, where the dimensionless operators are,
\begin{equation}
\hat{Q} = \frac{1}{\sqrt{2}}(\hat{a}^{\dagger} + \hat{a}), \quad \hat{P} = i\frac{1}{\sqrt{2}}(\hat{a}^{\dagger} - \hat{a})
\end{equation}
Let us now formulate the time evolution of such kicked system. Within one time-period, the system evolves freely under $\hat{H}_0$ followed by a kicking. The unitary time-evolution operator describing the dynamics of such system within a time period $T$ is \cite{Haake87},
\begin{equation}
\hat{\mathcal{F}} = e^{-i\hat{H}_0T} e^{-i\frac{\lambda _0}{\sqrt{S}}(\hat{a}^{\dagger} + \hat{a})\hat{S}_x}
\end{equation}
Within the Heisenberg picture any operator after the $n^{\rm th}$ kick can be obtained from, $\hat{O}^{(n)} = \hat{\mathcal{F}}^{\dagger} \hat{O}^{(n-1)} \hat{\mathcal{F}} = \hat{\mathcal{F}}^{\dagger ~ n} \hat{O} \hat{\mathcal{F}}^{n}$. Hence the stroboscopic time evolution of the relevant observables in different subspaces of the KDM namely, the spin operators $\hat{S}_{x,y,z}$, and the conjugate position (momenta) $\hat{Q}$ $(\hat{P})$ of an oscillator, can be written as a following matrix map,
\begin{equation}
\left[\hat{S}_{x,y,z}^{(n+1)}, \hat{Q}^{(n+1)}, \hat{P}^{(n+1)}\right]^{\rm T} = \hat{\mathcal{R}} \left[\hat{S}_{x,y,z}^{(n)}, \hat{Q}^{(n)}, \hat{P}^{(n)}\right]^{\rm T}
\label{Heisen_eqn}
\end{equation}
where, the time evolution matrix $\hat{\mathcal{R}}$ turns out to be,
\begin{equation}
\hat{\mathcal{R}}=\left(\begin{array}{ccccc}
\cos \tau & -\sin \tau & 0 & 0 & 0 \\
\cos \hat{z} \sin \tau & \cos \hat{z} \cos \tau & -\sin \hat{z} & 0 & 0 \\
\sin \hat{z} \sin \tau & \sin \hat{z} \cos \tau & -\cos \hat{z} & 0 & 0 \\
0 & 0 & 0 & \cos \bar{\tau} & \sin \bar{\tau} \\
-\bar{\lambda}_0 \cos \tau & -\bar{\lambda}_0 \sin \tau & 0 & -\sin \bar{\tau} & \cos \bar{\tau} \end{array}\right)
\label{R_matrix}
\end{equation}
with $\hat{z}=\bar{\lambda}_0 \left[\hat{Q}^{(n)}\cos \Omega T + \hat{P}^{(n)}\sin \Omega T \right]$, $\bar{\lambda}_0 = \sqrt{\frac{2}{S}}\lambda_0$, $\tau=\omega_0T$ and $\bar{\tau}=\Omega T$.
Finally, the average values of the time evolved operators are obtained by computing expectation w.r.t. the initially prepared state $\vert \psi_0\rangle$ of the system at $t=0$ i.e. $\langle \hat{O}\rangle = \langle \psi_0 \vert \hat{O} \vert \psi_0 \rangle$. Alternatively, in the Schr\"odinger picture, the state of the system is evolved under the kicked Dicke Hamiltonian in Eq.\ \eqref{Dicke_hamt}, and after the $n^{\rm th}$ kick, the time evolved state is obtained from $\vert \psi_n\rangle = \hat{\mathcal{F}}^n \vert \psi_0\rangle$. We would like to mention, due to unitarity of $\hat{\mathcal{F}}$, its eigenvalues have the form $e^{-i \phi_{\nu}}$ with eigenphase $\phi_{\nu}$ satisfying the eigenvalue equation, $\hat{\mathcal{F}} \vert \Phi _{\nu} \rangle = e^{-i \phi_{\nu}} \vert \Phi _{\nu} \rangle$ where, $\vert \Phi_{\nu} \rangle$ is the $\nu^{\rm th}$ Floquet state. Thus, using the spectral decomposition of the wavefunction in the eigenbasis of $\hat{\mathcal{F}}$, $\vert \psi_n\rangle$ can be expanded as,
\begin{equation}
\vert \psi_n \rangle = \sum _{\nu} c_{\nu} e^{-i \phi_{\nu} n} \vert \Phi _{\nu} \rangle,~ c_{\nu} = \langle \Phi_{\nu} \vert \psi_0 \rangle
\end{equation}
Therefore, at any instant $t = nT$, the expectation value of an observable $\hat{O}$ can be obtained from $\langle \hat{O} \rangle = \langle \psi_n \vert \hat{O} \vert \psi_n \rangle$.

\begin{figure}[t]
\begin{center}
\includegraphics[clip=true, width=1\columnwidth]{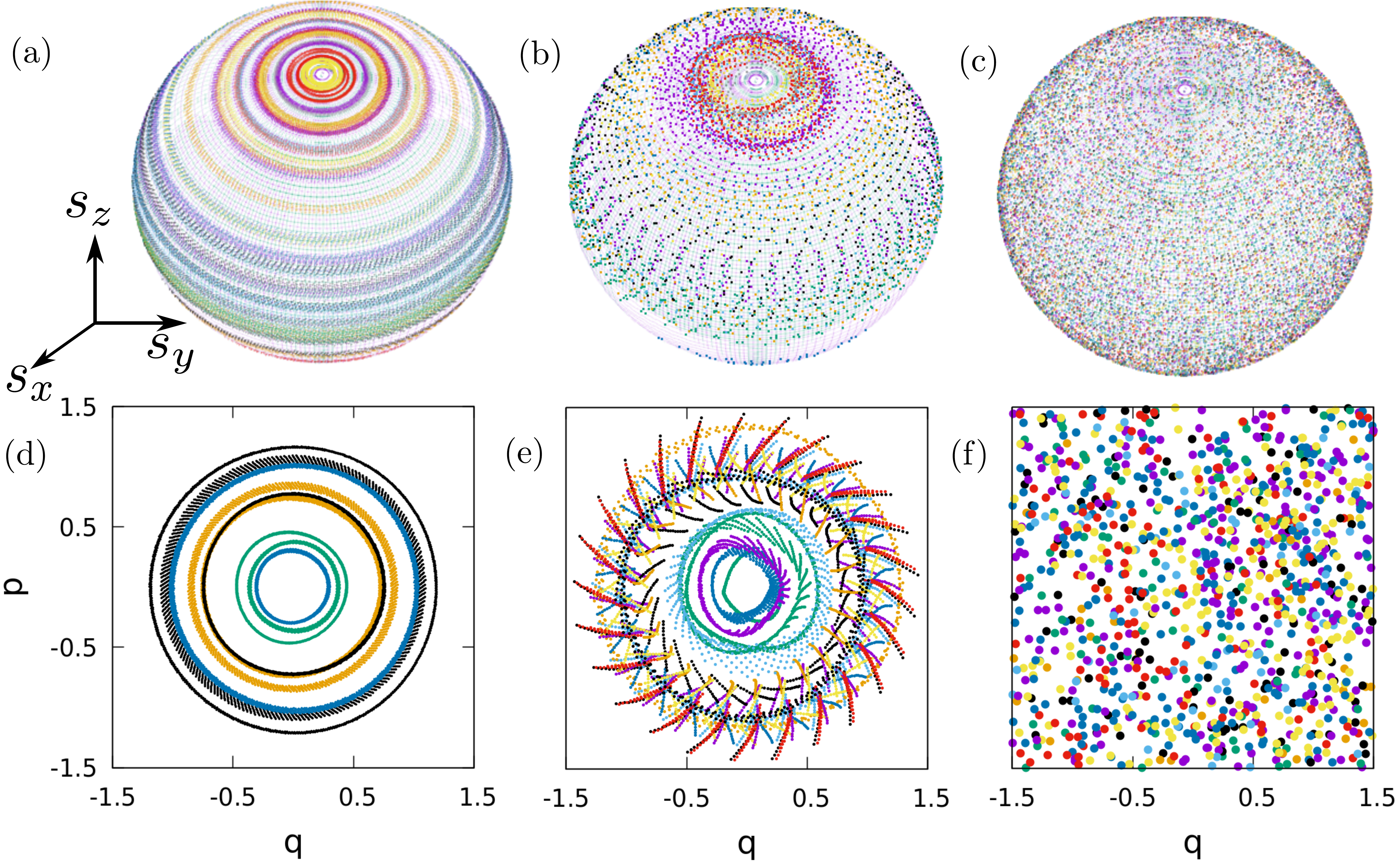}
\end{center}
\caption{{\it Phase portraits:} Stroboscopic plots on the Bloch sphere (top panel) and in the oscillator phase space (bottom panel) with increasing kicking strength $\lambda _0 = 0.01,~ 0.05,~ 0.6$ (from left to right). Different colors distinguish the trajectories starting from different initial conditions which are chosen randomly from a uniform distribution. We set $\Omega$ = 0.5, $T$ = 1.0. Here and in rest of the paper we will consider $\omega_0=1$ i.e. all the energy and time are measured in the unit of $\omega_0$ and $1/\omega_0$ respectively, e.g. $\Omega \equiv \Omega/\omega_0$, $\lambda_0 \equiv \lambda_0/\omega_0$ and $T \equiv \omega_0T$. Also, the $n^{\rm th}$ stroboscopic step corresponds to a time scale $t=nT$.}
\label{Phase_Port}
\end{figure}  

Having defined the general procedure to perform stroboscopic time evolution of the operators, we are now ready to define the out-of-time-order correlator (OTOC), which is of main interest in the present study,
\begin{eqnarray}
C(n) = \langle[\hat{W}(n),\hat{V}(0)]^{\dagger}[\hat{W}(n),\hat{V}(0)]\rangle
\label{correq1}
\end{eqnarray}
where, the expectation $\langle .\rangle$ is calculated either w.r.t. the initial state $\vert \psi(0)\rangle$, or using a density matrix $\hat{\rho}$ i.e. $\langle .\rangle\equiv{\rm Tr}(\hat{\rho}.)$. It can be noted that the OTOC $C(n)$ is related to the four-point correlation function $F(n)=\langle \hat{W}^{\dagger}(p) \hat{V}^{\dagger}(0) \hat{W}(p) \hat{V}(0) \rangle$, which is used to quantify the butterfly effect, and if $\hat{W}$ and $\hat{V}$ are unitary, the relation takes a simple form, $C(n) = 2(1-\mathrm{Re}[F(n)])$. In the present case, the operators $\hat{W}$ and $\hat{V}$ are chosen accordingly depending on bosonic or spin subspaces of the KDM. 

In general, the time evolution of $C(n)$ can have different dynamical regimes. During the initial period of scrambling, the dynamics of OTOC is governed by an interplay between the exponential growth of $C(n)$ and its diffusive counterpart, which depends on the complexity of an interacting system \cite{Swingle16, Patel17}. The resulting velocity with which a local perturbation propagates in space is known as {\it butterfly velocity}, analogous to the butterfly effect in a classically chaotic system. 
In the KDM, the growth of $C(n)$ can be observed by choosing the canonically conjugate position $\hat{Q} \equiv \hat{W}$ and momentum $\hat{P} \equiv \hat{V}$ operators of the oscillator subspace. Thus, in the semiclassical limit, representing these operators by the classical position $q$ and momentum $p$ variables, followed by replacing the commutator $[\hat{Q}(n),\hat{P}]$ with the Poisson bracket $i\hbar_{\rm eff}\{q(n),p\}$, Eq.\ \eqref{correq1} delivers, $C(n) = \hbar_{\rm eff}^2 \left(\partial q(n)/\partial q(0)\right)^2$. Therefore, for a system exhibiting chaotic dynamics, it is expected that $C(n)$ will grow as $\sim \hbar_{\rm eff}^2 e^{2\Lambda n}$, allowing us to extract the Lyapunov exponent $\Lambda$ from the growth rate of $C(n)$ within the Ehrenfest time $t_{\rm E} \sim 1/\Lambda\log (1/\hbar_{\rm eff})$ \cite{Beenakker02}. We will see in Sec.\ \ref{recap} that in our case, $\hbar_{\rm eff}=1/S$ plays the role of an effective Planck constant. 

However, for a generic quantum system with finite Hilbert space dimension, $C(n)$ cannot grow indefinitely, and exhibits a saturation after a long time evolution \cite{Garcia18, RayOTOC18,Arul20}. Moreover, not all the quantum systems have such a semiclassical limit which allows us to correspond the growth of $C(n)$ with the Lyapunov exponent. We thereby propose for an alternate diagnostic tool, which is the saturation value of $C(n)$ after a long time evolution, to capture the onset of chaos in a generic interacting quantum system. We will demonstrate this in the spin sector of the KDM by constructing the OTOC from one of the spin operators say $\hat{S}_z$ i.e. $C(n)=\langle [\hat{S}_z(n),\hat{S}_z(0)]^2\rangle$, followed by extracting its asymptotic value after the saturation of $C(n)$. However, before discussing what happens in these OTOC dynamics in different subspaces and how it indicates the onset of chaos, let us first revisit the classical counterpart of KDM followed by its correspondence with the spectral statistics.   

\section{Signature of chaos in KDM}
\label{recap}

In this section, we review the onset of chaos in the classical phase space dynamics, as well its reflection in the spectral statistics of the KDM, which has already been discussed elaborately in Ref.\ \cite{Ray16}, in the context of thermalization of such driven interacting system and its connection with the underlying chaotic dynamics.
The purpose of the present analysis is two fold. First, it is relevant for the direct investigation of local chaos and the degree of ergodicity via the out-of-time-order correlator (OTOC) study (cf. Sec.\ \ref{OTOC}). Secondly, the analysis of fixed point structure in classical dynamics is important for identifying the effect of scarring in such system, which will be presented in Sec. \ \ref{scar}.

\begin{figure}[t]
\begin{center}
\includegraphics[clip=true, width=1\columnwidth]{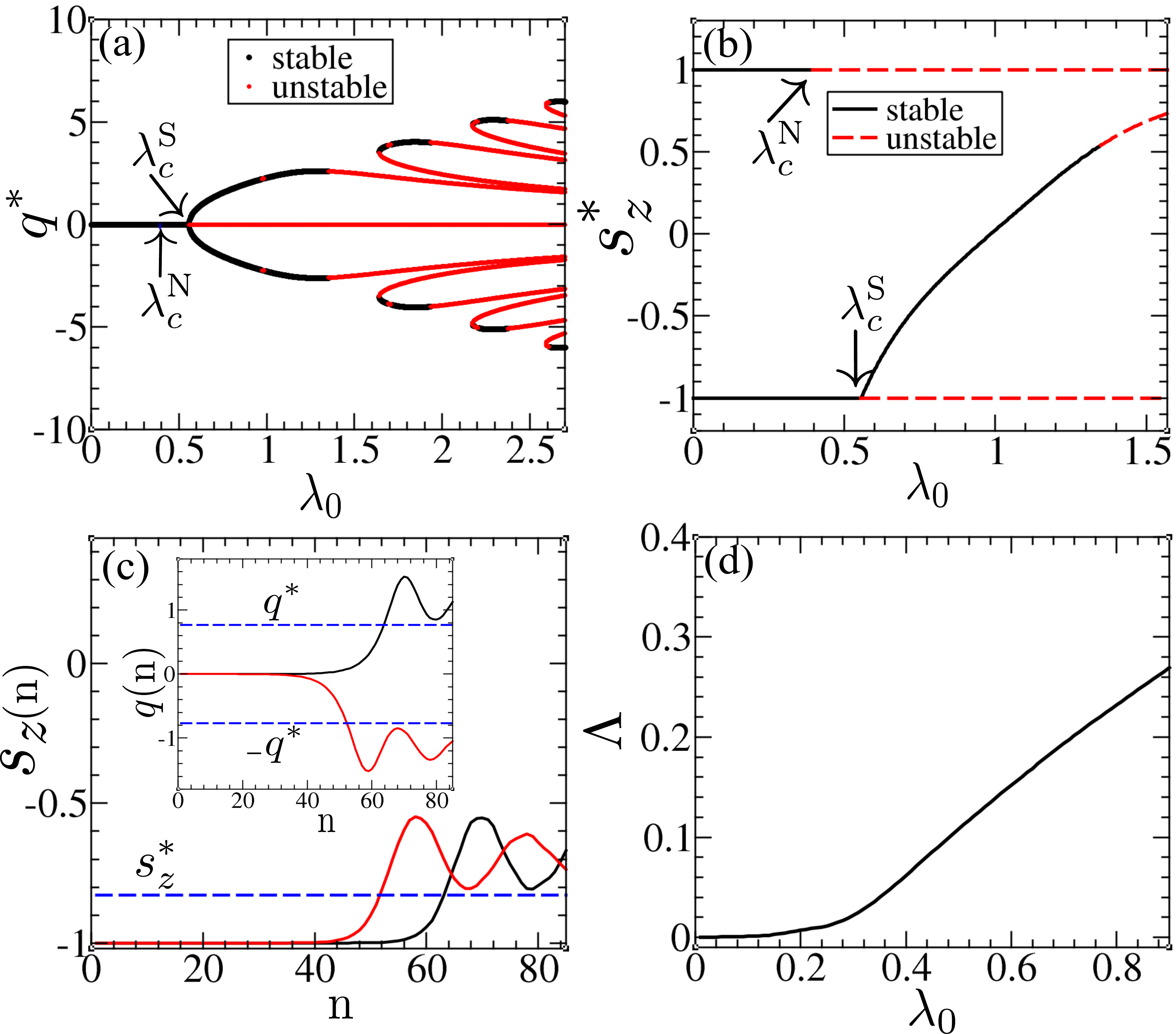}
\end{center}
\caption{{\it Fixed point bifurcation:} Fixed points (a) $q^*$ and (b) $s^{*}_{z}$ are plotted as a function of $\lambda_0$. The stable (unstable) branches are represented by black (red) lines. The instability of the trivial fixed points $s_{\pm}^*$ are marked by the arrowheads. (c) Classical dynamics of $s_z$ ($q$ in the inset) starting from the fixed point $s_{-}^*$ with a small perturbation given in terms of the initial polar angle $\theta = \pi \pm \epsilon$ ($\epsilon<<1$), shown in black and red lines respectively at $\lambda_0=0.6$, just after it becomes unstable at $\lambda^{\rm S}_c = 0.556$. The blue-dashed horizontal lines denote the non-trivial fixed points $s^{*}_z$ ($\pm q^{*}$ in the inset). (d) The maximum Lyapunov exponent $\Lambda$ is plotted as a function of $\lambda_0$. Other parameters are $\Omega$ = 0.5 and $T$ = $\pi/3$, which will be kept same for rest of the figures unless mentioned otherwise.}
\label{Stability}
\end{figure}

\subsection{Classical dynamics}

The classical counterpart of the KDM can be analyzed by suitably taking the classical limit of the Heisenberg equations in Eq.\ \eqref{Heisen_eqn}. In order to do that, we re-define the spin and bosonic operators scaled by the spin magnitude $S$ in the following way: $\hat{s}_i=\hat{S}_i/S$, $\hat{q}=\hat{Q}/\sqrt{S}$ and $\hat{p}=\hat{P}/\sqrt{S}$. Therefore, the commutators of spin and bosonic operators can be re-written as, $\left[\hat{s}_i,\hat{s}_j\right]=i\hbar_{\rm eff}\epsilon _{ijk}\hat{s}_k$ and $\left[\hat{q},\hat{p}\right]=i\hbar_{\rm eff}$, where $\hbar_{\rm eff} = 1/S$ becomes the effective Planck's constant. It is now easy to see that, in the limit $S \rightarrow \infty$, or equivalently $\hbar_{\rm eff} \rightarrow 0$, the commutators vanish and thus the quantum operators can be treated as classical variables such as, $q(p)$ are the canonically conjugate position (momenta) of the oscillator and $s_i$'s represent the spin components of a spin vector $\vec{s} \equiv (s_x, s_y, s_z)= (\sin\theta \cos\phi, \sin\theta \sin\phi, \cos\theta)$, where $\theta$ and $\phi$ are the polar and the azimuthal angles respectively, representing the orientation of $\vec{s}$ on a Bloch sphere. The resulting stroboscopic time evolution of the classical dynamical variables are described by the following Hamiltonian map,
\begin{equation}
\left[s_{x,y,z}^{(n+1)}, q^{(n+1)}, p^{(n+1)}\right]^{\rm T} = \mathcal{R} \left[s_{x,y,z}^{(n)}, q^{(n)}, p^{(n)}\right]^{\rm T}
\label{Ham_map}
\end{equation}
where, $\mathcal{R}$ is defined in Eq.\ \eqref{R_matrix} with the exception that all the operators are now replaced by the c-number classical variables and the coupling $\bar{\lambda}_0$ becomes $\sqrt{2}\lambda_0$. Therefore, the Hamiltonian map in Eq.\ \eqref{Ham_map} is independent of the spin magnitude $S$ and satisfy the conservation of the total spin i.e. $s_x^2 + s_y^2 + s_z^2 = 1$. In Fig.\ \ref{Phase_Port}, we have shown the phase portraits by iterating the Hamiltonian map, both in the oscillator space (constructed by canonically conjugate $q$ and $p$ variables) and on the Bloch sphere. While for small kicking strength $\lambda_0$, the phase space trajectories remain regular, it undergoes a chaotic phase space dynamics in both the subspaces with increasing $\lambda_0$.

To analyze the onset of chaos demonstrated in Fig.\ \ref{Phase_Port}, let us first look into the fixed points of the Hamiltonian map which can be obtained from, $s_{i}^{(n+1)} = s_{i}^{(n)} = s_i^*$, $q^{(n+1)} = q^{(n)} = q^*$ and $p^{(n+1)} = p^{(n)} = p^*$. This delivers a set of trivial fixed points: $s^*_{\pm} \equiv \{s_x^*, s_y^*, s_z^*, q^*, p^*\} = \{0,0,\pm 1,0,0\}$, which represent the excited and ground states of the Dicke model in absence of any kicking. Moreover, the fixed point $s^*_{+}$ is closely related to the `$\pi$-oscillation' in Bose-Josephson junction \cite{Smerzi97, Oberthaler10, Sinha19}. Further non-trivial fixed points are obtained by solving the following non-linear equation for $q^*$,
\begin{equation}
\frac{(q^*)^2 \tan^2\left(\frac{\Omega T}{2}\right)}{\cos^2\left(\frac{T}{2}\right)} \left[1 + \cot^2\left(\frac{\lambda _0 q^*}{\sqrt{2}}\right) \sin^2\left(\frac{T}{2}\right) \right] = \lambda _0^2/2. 
\label{Q_c_sol}
\end{equation}
The non-zero values of the other variables corresponding to this fixed point can be easily obtained from the Hamiltonian map with the fixed point conditions. Stability of these fixed points is analyzed from the eigenvalues of the Jacobian, $J_{ij} = \partial A_i^{(n+1)}/\partial A_j^{(n)}$ with $\{A_i\} \equiv \{s_x, s_y, s_z, q, p\}$. Stability is ensured if the magnitudes of all the eigenvalues are unity \cite{Haake, Strogatz}. Let us now look into the instability of the two trivial fixed points $s^*_{\pm}$, which remain dynamically stable for smaller kicking strength $\lambda_0$. Amongst them, $s^*_{+}$ loses the stability first, as $\lambda_0$ is increased above a critical value given by, 
\begin{equation}
\lambda_c^{\rm N} = \sqrt{(\cos T - \cos \Omega T)^2/(2\sin T\sin \Omega T)}
\end{equation}
Further increase in $\lambda_0$ destabilizes the other trivial fixed point $s^*_{-}$ corresponding to the south pole at a critical kicking strength $\lambda_c$, where the first pair of non-trivial fixed points appear. By solving Eq.\ \eqref{Q_c_sol}, we obtain an expression for $\lambda_c$ given by,
\begin{equation}
\lambda_c^{\rm S} = \sqrt{2\tan(\Omega T/2)\tan(T/2)}.
\label{l_c}
\end{equation} 
It can be noted that, in the limit $T\rightarrow0$, one can obtain the Dicke model as the time averaged Hamiltonian with an effective coupling $\lambda_{\rm eff}=\lambda_0/T$. The bifurcation of $q^{*}$ at $\lambda^{\rm S}_{c}$ [see Fig.\ \ref{Stability}a] corresponds to the `superradiant' phase transition in Dicke model as $q^{*}$ becomes non-zero. 
However, in presence of kicking, the stable non-trivial fixed points obtained after bifurcation become unstable with increasing $\lambda_0$ [see Fig.\ \ref{Stability}(a,b)]. In Addition, the fixed point bifurcation is only associated with the south pole of the Bloch sphere; it does not occur in the north pole as clearly shown in Fig.\ \ref{Stability}b. Its consequence in the dynamics is demonstrated in Fig.\ \ref{Stability}c, where an initial point chosen very close to the south pole $(s^*_-)$, tends to move towards the non-trivial stable fixed points. We will later see that even after becoming unstable, the fixed points $s^*_{\pm}$ retain its trace in the corresponding quantum dynamics and does not become ergodic, which we will relate to effect of quantum scar. Further increase in $\lambda_0$ results in a proliferation of fixed points, and their stability region gradually decreases leading to a completely chaotic phase space, which is discussed elaborately in Ref.\ \cite{Ray16}.   

We complete this section by computing the Lyapunov exponent (LE), which is a standard measure for degree of phase space chaos. This is done by a QR decomposition of the Jacobian $J$ obtained by linearizing the Hamiltonian map given in Eq.\ \eqref{Ham_map} \cite{Lauterborn90}. In Fig.\ \ref{Stability}d, we have plotted the maximum LE $\Lambda$, which is averaged uniformly over the phase space, as a function of $\lambda_0$. As expected, for small kicking strength, $\Lambda$ remains vanishingly small, and starts rising around $\lambda_0 \sim \lambda_c^{\rm N}$, gradually increasing as $\lambda_0$ is increased further. This reflects the onset of average chaos observed in the phase space of KDM. We will later correspond the classical LE obtained in this way with the growth rate of out-of-time-order correlator as its quantum counterpart.   

\begin{figure}[t]
\begin{center}
\includegraphics[clip=true,width=1\columnwidth]{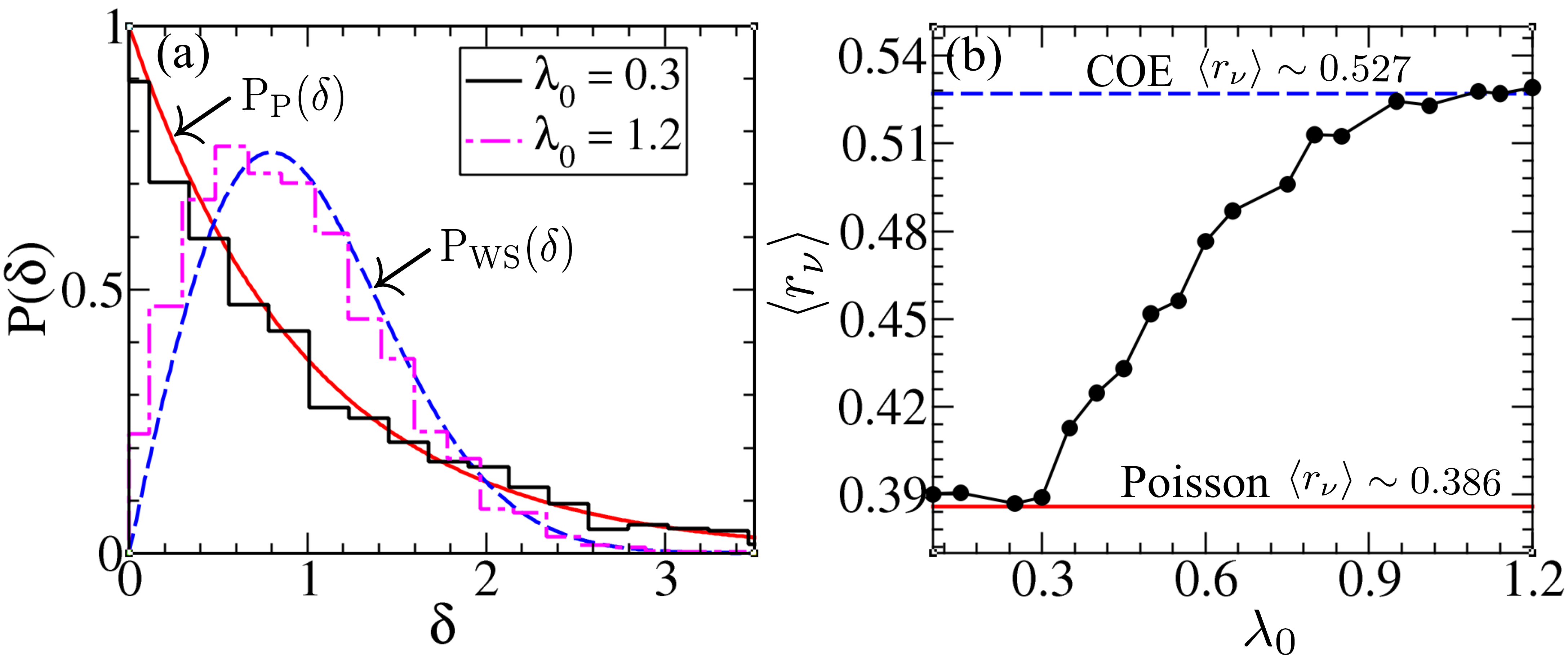}
\end{center}
\caption{{\it Spectral statistics:} (a) Histograms of the eigenphase spacings, ${\rm P(\delta)}$ are plotted for $\lambda _0=0.3$ (solid black line) and $\lambda _0=1.2$ (pink dashed-dotted line), which is compared with the distributions ${\rm P}_{\rm P}(\delta)$ (solid red) and ${\rm P}_{\rm WS}(\delta)$ (blue dashed) respectively. (b) Variation of average level spacing ratio $\langle r_{\nu} \rangle$ is shown with increasing kicking strength $\lambda_0$. The solid red and blue dashed lines mark its values for Poisson $(\sim 0.386)$ and circular orthogonal ensemble of RMT $(\sim 0.527)$ respectively. We have considered the even symmetry sector of $\hat{\mathcal{F}}$ to generate these plots which has no qualitative difference from the odd sector. For numerical computations shown here and in rest of the paper, we set the spin magnitude $S = 20$ and truncate the bosonic Hilbert space at $N_{\rm max} = 120$.}
\label{lev_spacing}
\end{figure}

\subsection{Spectral Statistics}
\label{Lev_stat}

An alternate way to characterize the two broad class of dynamics, regular and chaotic, is the spectral statistics of the corresponding quantum Hamiltonian. While according to the Berry-Tabor conjecture, Poisson distribution of the energy level spacing implies a regular phase-space dynamics \cite{Berry77}, on the other hand, BGS conjecture says that, Wigner-Dyson distribution reflecting the repulsion between energy levels is an indicator of the onset of chaos in the classical dynamics \cite{Bohigas84}.
Analogously, in a periodically driven quantum system, the spectral statistics of the Floquet operator can be analyzed from its quasi-energies \cite{Haake,Haake87}.
In this subsection, we study the fingerprint of chaos in the kicked Dicke model based on the statistical properties of the eigenphases $\phi_{\nu}$ of $\hat{\mathcal{F}}$. 

Before that, we first recall that Dicke model has a conserved parity corresponding to the parity operator $\hat{\Pi} = e^{i \pi \hat{N}}$ \cite{Brandes03}, where $\hat{N} = \hat{a}^{\dagger}\hat{a} + \hat{S}_z + S$. The operator $\hat{\Pi}$ has two eigenvalues $\mathcal{P} = \pm 1$, that depends on the odd or even integer eigenvalues of $\hat{N}$. The parity conservation also holds in the Floquet dynamics. Based on the values of $\mathcal{P}$, we separate out the Floquet eigenmodes into even $(\mathcal{P} = 1)$ and odd $(\mathcal{P} = -1)$ parity sectors. The statistical analysis is performed separately for the two symmetry sectors and we did not find any qualitative difference between them. Therefore, we present the results on spectral statistics by considering the even symmetry sector without any loss of generality.

In order to compute the distribution ${\rm P}(\delta)$ of the spacings between the eigenphases, $\delta_{\nu} = \phi_{\nu + 1} - \phi_{\nu}$, we follow the standard procedure as outlined in \cite{Haake} so as to preserve the normalization and mean value to unity. The resulting level spacing distribution is plotted in Fig.\ \ref{lev_spacing}a. We clearly observe that ${\rm P}(\delta)$ follows Poisson statistics given by ${\rm P}_{\rm P}(\delta) = e^{-\delta}$ for small $\lambda _0$, where the underlying phase space is regular. On the other hand, as $\lambda_0$ is increased, the spacing distribution ${\rm P}(\delta)$ exhibits level repulsion and agrees well with the orthogonal ensemble of RMT described by the Wigner-Surmise, ${\rm P}_{\rm WS}(\delta) = (\pi \delta/2)~e^{-\pi \delta ^2/4}$ in the deep chaotic regime. Although, the eigenphases are distributed over a circle in this regime and the spacing distribution should follow circular orthogonal ensemble (COE) of RMT, it is however, almost indistinguishable from the Gaussian orthogonal ensemble (GOE) of RMT \cite{Rigol14, Haake}, which we have used for a comparison with our numerical results.

To interpolate between the two extreme ends, namely, the crossover from Poisson to the orthogonal class of RMT reflecting a change from regular to chaotic dynamics by tuning $\lambda_0$, we compute the average level spacing ratio $\langle r_{\nu}\rangle$ from the consecutive level spacings $\delta_{\nu}$, where the definition of $r_{\nu}$ follows \cite{Huse13,Bogomolny13},
\begin{equation}
r_{\nu} = \frac{\text{min}(\delta _{\nu+1},\delta _{\nu})}{\text{max}(\delta _{\nu+1},\delta _{\nu})}
\label{r_avg}
\end{equation}
In Fig.\ \ref{lev_spacing}b, we have plotted $\langle r_{\nu} \rangle$  as a function of $\lambda_0$, where $\langle r_{\nu} \rangle$ denotes an average of $r_{\nu}$ over all the eigenmodes belonging to a given symmetry sector of $\mathcal{\hat{F}}$. Clearly the variation ranges between two limits: Poisson statistics with $\langle r_{\nu} \rangle = 2 \ln 2 - 1 \approx 0.386$ as lower limit, and circular orthogonal ensemble of RMT, $\langle r_{\nu} \rangle \approx 0.527$ \cite{Rigol14} as the upper limit which are marked by the horizontal lines in Fig.\ \ref{lev_spacing}b. It is noteworthy that the upturn of $\langle r_{\nu} \rangle$ with increasing $\lambda_0$ coincides with that of average Lyapunov exponent in Fig.\ \ref{Stability}d, thereby capturing the change in dynamical behavior as well as the onset of chaos with increasing kicking strength. 

However, a direct estimation of the degree of chaos in the quantum dynamics of our driven interacting system is still missing. In the next section, we aim to investigate this missing link from the dynamics of out-of-time-order correlators. 
 
\section{OTOC dynamics}
\label{OTOC}

In this section, we propose an alternate method to study the onset of chaos in the KDM using the stroboscopic dynamics of out-of-time-order correlator $C(n)$ defined in Sec.\ \ref{model}. To probe the local chaotic behavior in the phase space, $C(n)$ is evaluated by using the coherent state representing a phase space point semiclassically. For Dicke model, such a coherent state would be a product state given by,
\begin{equation}
\vert \alpha ; \theta, \phi \rangle = \vert \alpha \rangle \otimes \vert \theta, \phi \rangle
\end{equation}
The states $\vert \alpha \rangle$ and $\vert \theta, \phi \rangle$ represent the coherent states of a harmonic oscillator \cite{sudarsan} and of a spin of magnitude $S$ \cite{Radcliff} respectively. They can be written as follows,
\begin{equation}
\vert \alpha \rangle = e^{-\vert \alpha \vert ^2/2} e^{\alpha \hat{a}^{\dagger}}\vert 0 \rangle,~ \vert \theta, \phi \rangle = (1 + \vert z \vert ^2)^{-S} e^{z \hat{S}_{+}} \vert S,-S \rangle
\label{coherent}
\end{equation}
where the c-numbers, $\alpha$ and $z$ represent the classical phase space variables of an oscillator, $\alpha = (q+ip)/\sqrt{2}$ and a spin, $z = e^{-i \phi} \tan \theta/2$ respectively.
The evaluation of OTOC $C(n)$ is done by choosing the appropriate operators for the two subspaces. In case of an oscillator, the natural choice are the canonically conjugate position $\hat{W}=\hat{Q}/\sqrt{S}=\hat{q}$ and momentum $\hat{V}=\hat{P}/\sqrt{S}=\hat{p}$ operators for evaluation of the correlator $C_{qp}(n)$. Its growth rate in the chaotic regime is expected to capture the Lyapunov exponent in the semiclassical limit (cf. Sec.\ \ref{model}). Hence, the degree of chaos can be studied from a change in the growth rate of $C_{qp}(n)$ by tuning the kicking strength $\lambda_0$. On the other hand, in the spin subspace, we set $\hat{W}=\hat{V}=\hat{S}_{z}/S=\hat{s}_z$, in order to evaluate the  correlator $C_{S_z}(n)$. We expect that for small $\lambda_0$, the correlator $C_{S_z}$ becomes vanishingly small as the operator $\hat{S}_z$ commutes with the KDM Hamiltonian given in Eq.\ \eqref{Dicke_hamt}, and as $\lambda_0$ is increased, the spreading of spin operators can result in a non vanishing value of $C_{S_z}(n)$, which can be useful for identifying the onset of chaos.

\begin{figure}[t]
\begin{center}
\includegraphics[clip=true,width=1\columnwidth]{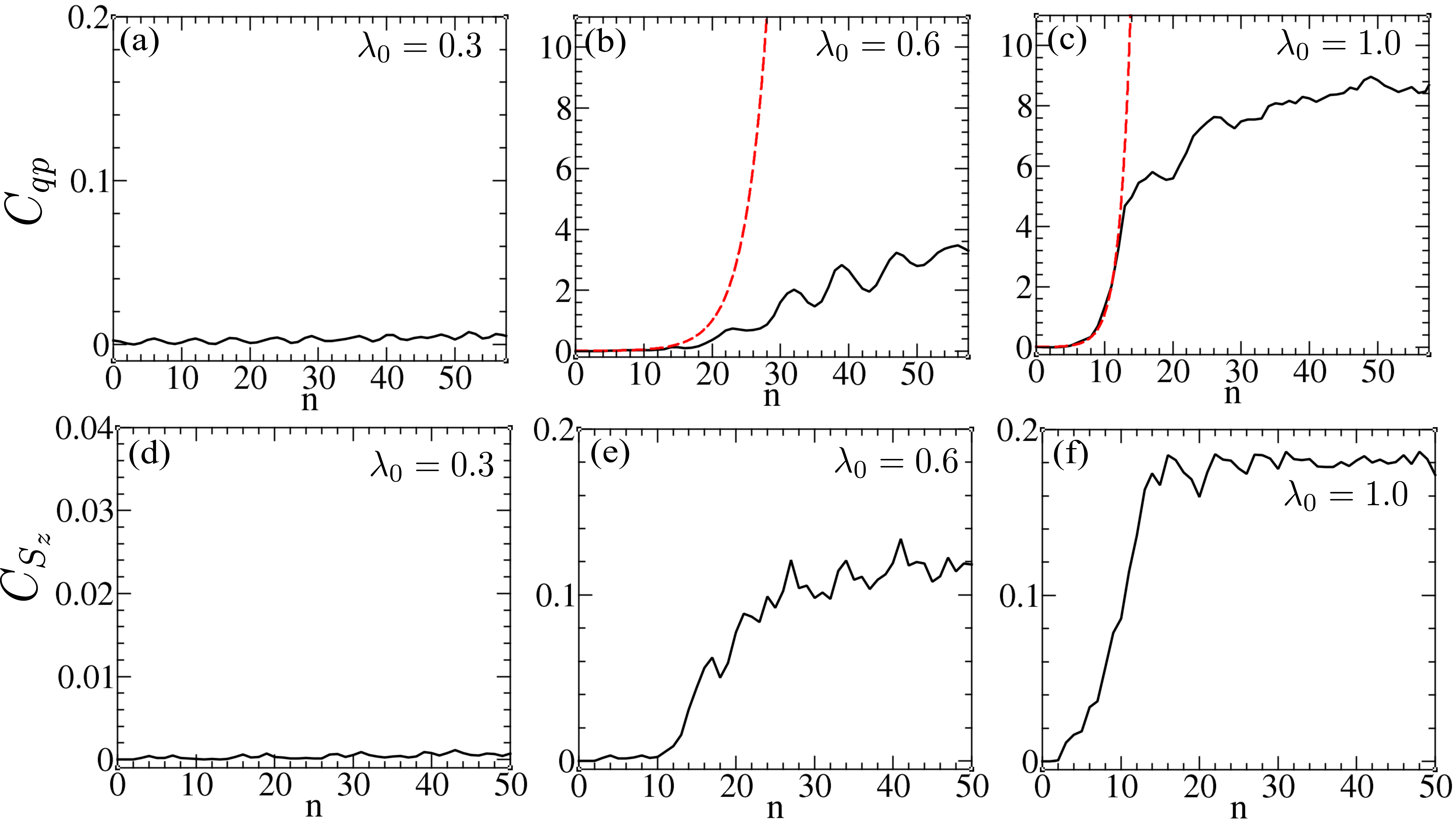}
\end{center}
\caption{{\it Dynamics of OTOC:}  Stroboscopic dynamics of $C_{qp}(n)$ (top panel) and $C_{S_z}(n)$ (bottom panel) plotted with increasing $\lambda_0$ (from left to right), computed using a coherent state which represents a typical point in the classical phase space. The red dashed lines in (b) and (c) are generated from $\hbar_{\rm eff}^2e^{2\Lambda n}$, where $\Lambda$ is the classical LE for a given $\lambda_0$ (see Fig.\ \ref{Stability}d), and $\hbar_{\rm eff}=1/S$ is the effective Planck constant.}

\label{cn_dynamics}
\end{figure}

In Fig.\ \ref{cn_dynamics}, we have summarized the dynamical behavior of $C_{qp}(n)$ and $C_{S_z}(n)$, evaluated from a coherent state corresponding to an arbitrary phase space point, with increasing kicking strength $\lambda_0$. While for small $\lambda_0$, the growth of the correlators remain significantly small indicating a regular phase space dynamics, it increases with increasing $\lambda_0$, indicating the onset of chaos in both the subspaces. As the system approaches the chaotic regime for larger $\lambda_0$, the quantum-classical correspondence becomes weaker \cite{Ray16} due to a decrease in the Ehrenfest time, and as a consequence, we observe a sharp rise of the OTOCs within a first few drives which is clearly visible in Fig.\ \ref{cn_dynamics}(c,f). 
Such a growth of OTOC for the oscillator is expected to follow $\hbar^2_{\rm eff}e^{2\Lambda n}$ in the chaotic regime (cf. Sec.\ \ref{model}). For further illustration, this functional behavior is compared with the numerically evaluated $C_{qp}(n)$ for different $\lambda_0$ in Fig.\ \ref{cn_dynamics}(b,c). In the mixed phase space regime i.e. for an intermediate $\lambda_0$, where $\langle r_{\nu} \rangle$ lies in between Poisson and COE values, we observe a growth of $C_{qp}(n)$, however, the growth rate deviates from the Lyapunov exponent which is evident from Fig.\ \ref{cn_dynamics}b. In the deep chaotic regime for larger $\lambda_0$, as $\langle r_{\nu} \rangle$ approaches the COE value, the agreement between the growth rate of $C_{qp}(n)$ and the Lyapunov exponent becomes closer, which is shown in Fig.\ \ref{cn_dynamics}c. Such a behavior has also been observed in a non interacting chaotic system \cite{Galitski17}.

\begin{figure}[t]
\begin{center}
\includegraphics[clip=true,width=1\columnwidth]{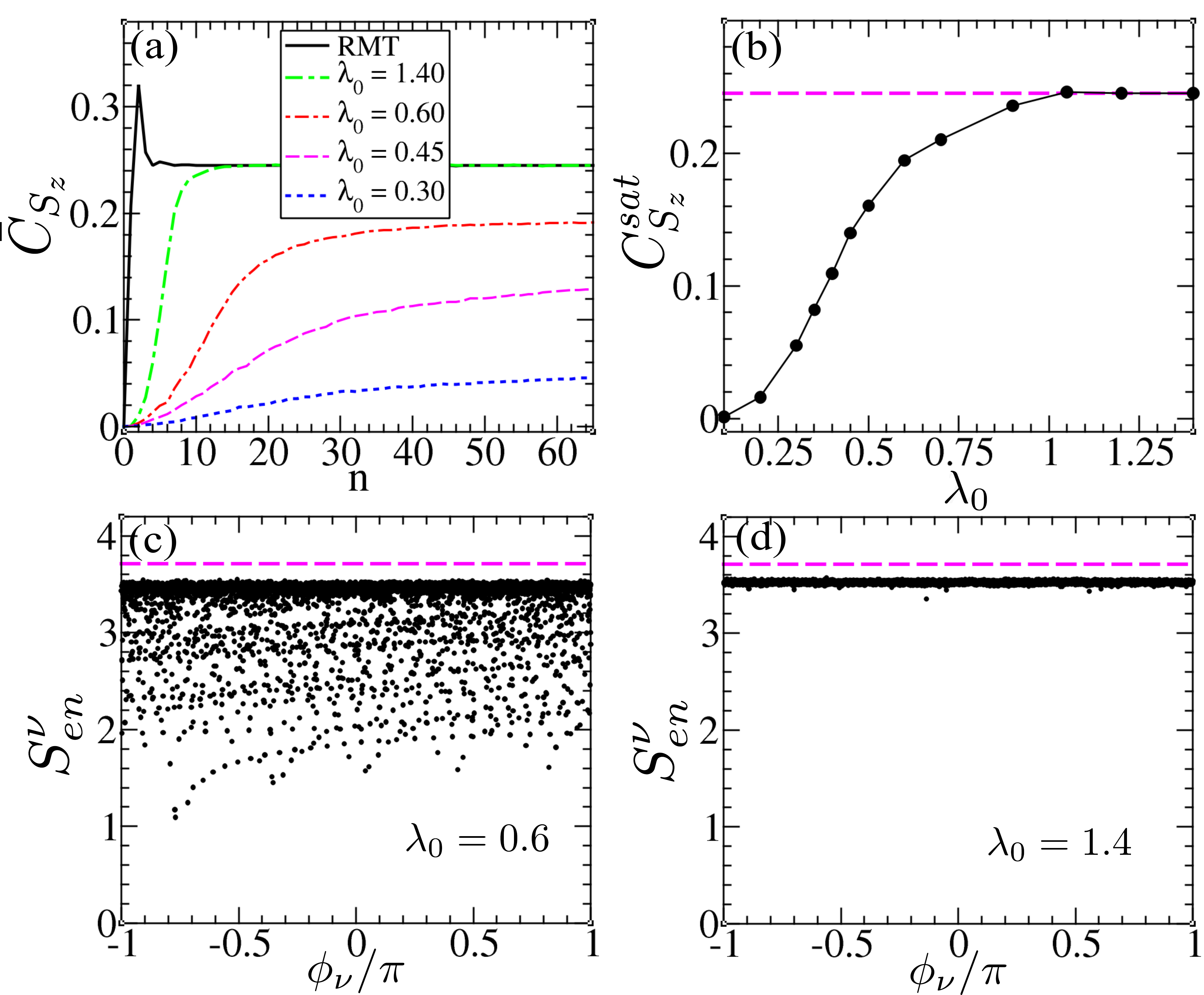}
\end{center}
\caption{{\it Saturation of OTOC and entanglement:} (a) Stroboscopic dynamics of $\bar{C}_{S_z}(n)$ for different kicking strengths $\lambda_0$. (b) Variation of long-time saturation $C_{S_z}^{sat}$ with increasing $\lambda_0$. The solid black line in (a) and the horizontal dashed line in (b) are obtained from time evolution under $\hat{\mathcal{U}}_{\rm GOE}$. (c,d) Entanglement entropy $S_{\rm en}^{\nu}$ computed from the eigenstates of $\hat{\mathcal{F}}$. The horizontal dashed line marks its maximum limit $\ln (2S+1)$.}

\label{cqq_sat}
\end{figure}

Let us now analyze the behavior of the correlator $C_{S_z}(n)$ in the spin subspace. From Fig.\ \ref{cn_dynamics}(d-f), we see that the growth and magnitude of OTOC $C_{S_z}(n)$ increase with increasing $\lambda_0$ exhibiting a similar behavior observed for $C_{qp}(n)$ in the oscillator subspace. However, we would like to point out that unlike $C_{qp}(n)$, the growth rate of $C_{S_z}(n)$ does not capture the Lyapunov exponent in the chaotic regime. Instead, it is noteworthy that the saturation value of $C_{S_z}(n)$ increases with increasing degree of chaos in the phase space, as the kicking strength $\lambda_0$ is increased. Here it is important to mention that, although $C_{qp}(n)$ also exhibits a saturation, but it has a strong dependence upon the numerical truncation effect in the bosonic sector. Ideally, it grows without any saturation in the unstable/chaotic regime \cite{Biswas20}. Therefore, to investigate how the overall chaotic behavior is reflected from the saturation value of OTOC, we focus on the following spin correlator,
\begin{equation}
\bar{C}_{S_z}(n) = -{\rm Tr} \left(\hat{\rho}_{\rm mc} \left[\hat{s}_z(n),\hat{s}_z(0)\right]^2\right),
\label{OTOC_sx}
\end{equation}  
which is evaluated using the microcanonical density matrix $\hat{\rho}_{\rm mc}= \sum_{\nu} \vert \Phi_{\nu} \rangle \langle \Phi_{\nu} \vert/\mathcal{N}$. The stroboscopic time evolution of $\bar{C}_{S_z}(n)$ and its long time saturation value $C_{S_z}^{sat} \equiv \bar{C}_{S_z}(n\rightarrow \infty)$ are shown in Fig.\ \ref{cqq_sat}(a,b) with increasing kicking strength $\lambda_0$. While in the classically regular regime, $\bar{C}_{S_z}(n)$ does not grow and remains vanishingly small, it increases for larger kicking strength. Clearly, the crossover from regular to chaotic dynamics with increasing $\lambda_0$ is captured by the increase in the saturation value $C_{S_z}^{sat}$ of the correlator, which finally converges to a maximum limit in the ergodic regime, denoted by $C^{\rm max}_{S_z}$, as depicted in Fig.\ \ref{cqq_sat}b. To estimate the upper limit of $C_{S_z}^{sat}$ in the ergodic regime, we assume an approximate form of the Floquet matrix given by, $\hat{\mathcal{U}}_{\rm GOE}=e^{i\hat{H}_{\rm GOE}T}$, where $\hat{H}_{\rm GOE}$ is a random matrix of GOE class. The saturation value of OTOC $\bar{C}_{S_z}(n)$ obtained using such random matrix ensemble is marked by a horizontal dashed line in Fig.\ \ref{cqq_sat}b, which quite remarkably agrees with the observed upper limit $C^{\rm max}_{S_z}$ attained at larger $\lambda_0$. It turns out that in this ergodic regime, for sufficiently large value of $S$, the saturation value approaches to $C^{\rm max}_{S_z} \simeq 2 \langle \hat{s}^2_z \rangle^2_{\rm mc}$, which is also in accordance with ETH, where $\langle . \rangle_{\rm mc}$ represents the microcanonical average.

An alternate way to understand the ergodic properties of a driven quantum system relies on the entanglement entropy of the Floquet eigenvectors $\vert \Phi_{\nu}\rangle$. By tracing out the bosonic degrees of freedom, the entanglement entropy of the spin sector can be written as,
\begin{equation}
S_{en}^{\nu} = -{\rm Tr}\hat{\rho_{\rm S}^{\nu}}\ln\hat{\rho_{\rm S}^{\nu}}, \quad \hat{\rho}_{\rm S}^{\nu} = {\rm Tr}_{\rm osc} \left( |\Phi_{\nu}\rangle \langle \Phi_{\nu}| \right),
\end{equation}
where $\hat{\rho}^{\nu}_S$ is the reduced spin density matrix. It is expected that, in the ergodic regime the entanglement entropy will approach to its maximum value $\ln (2S+1)$ corresponding to completely random eigenvectors \cite{page}. In Fig.\ \ref{cqq_sat}(c,d), we have plotted $S_{en}^{\nu}$ for different kicking strengths $\lambda_0$ and contrasted it with this maximum limit. For an intermediate $\lambda_0$ where a classically mixed phase space appears, as well $\langle r_{\nu} \rangle$ takes a value in between the two extremes, the entanglement entropy $S^{\nu}_{en}$ of a large number of Floquet eigenstates remain scattered below its maximum limit [see Fig.\ \ref{cqq_sat}c]. Whereas, with increasing $\lambda_0$, as the system enters into the deep chaotic regime, the entanglement entropy of the Floquet states form a band like structure close to its maximum limit which is clearly observed in Fig.\ \ref{cqq_sat}d. Such a behavior of the Floquet states in the deep chaotic regime justifies the approximate form of the Floquet matrix $\hat{\mathcal{U}}_{\rm GOE}$ which we have constructed using a random symmetric matrix $\hat{H}_{\rm GOE}$ as mentioned above. Furthermore, the change in spectral statistics from Poisson to an orthogonal class of RMT with increasing $\lambda_0$ indicates a crossover from regular to ergodic dynamics, and can be understood as an interplay between the two different classes of random matrices i.e Poisson and GOE \cite{Brody_rmp, Wettig12}. The behavior of OTOC as well as its saturation value using such a mixed random matrix model can effectively capture the onset of chaos in a system independent manner, which is presented in appendix\ \ref{Eff_RMT}.
 
\section{Effect of scarring}
\label{scar}

\begin{figure}[t]
\centering
\includegraphics[clip=true,width=1\columnwidth]{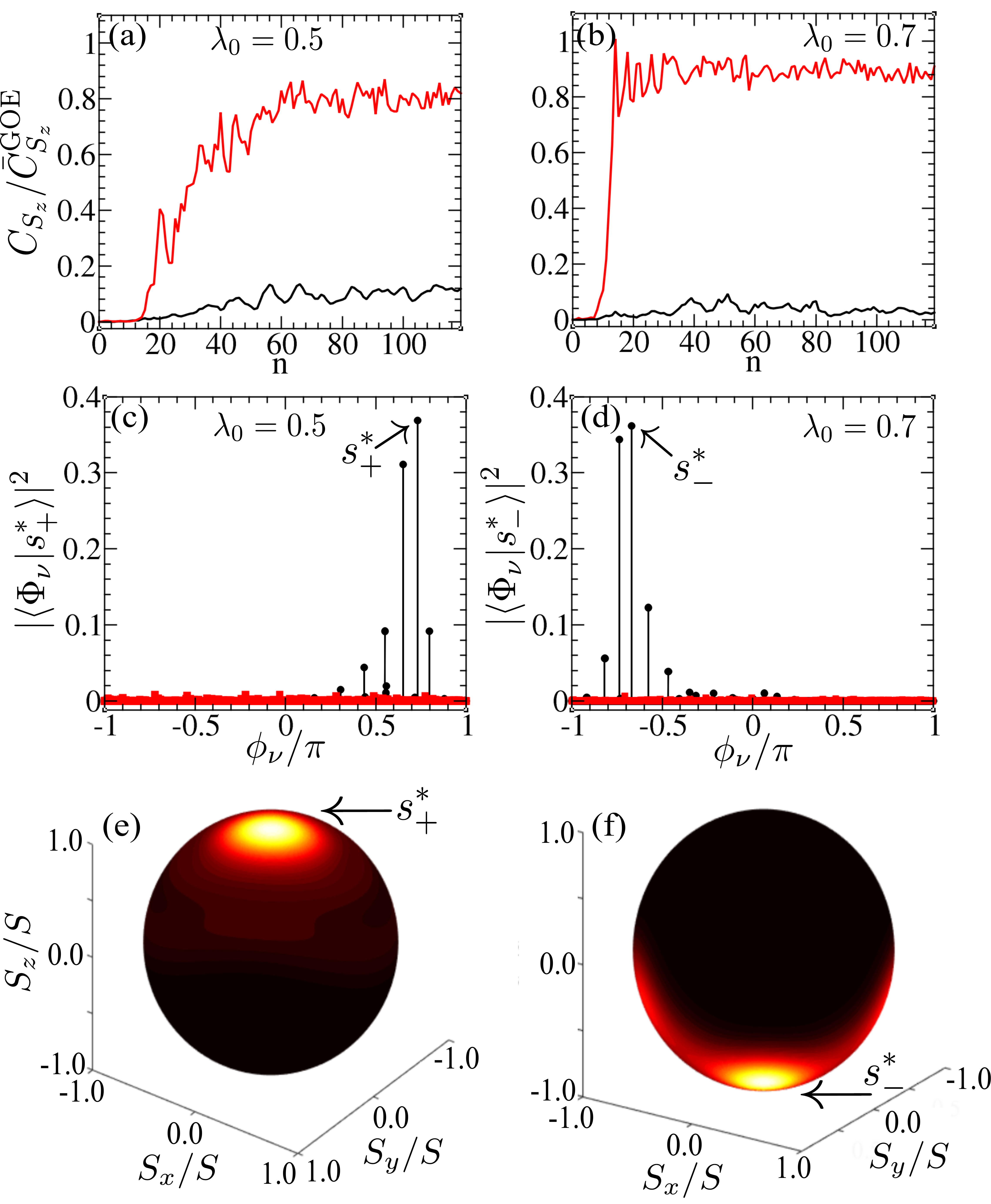}
\caption{{\it Effect of scarring:} (a,b) Stroboscopic dynamics of the scaled correlator $C_{S_z}(n)/\bar{C}^{\rm GOE}_{S_z}$ evaluated using a coherent state $\vert s_{\pm}^* \rangle$ (solid black). (c,d) Overlap of $\vert s_{\pm}^* \rangle$ with the Floquet states $\vert \Phi_{\nu}\rangle$ (black circles). (e,f) Husimi distribution $Q(\theta,\phi)$ of the Floquet state having maximum overlap, marked by arrowheads in (c,d) are plotted on the Bloch sphere. The results in (a-d) obtained by using $\vert s_{\pm}^* \rangle$ (shown in black) are contrasted with those obtained from a typical point in a chaotic region of the phase space (shown in red). The left (right) column corresponds to the fixed points $s_{\pm}^*$ and $\lambda_0$ is chosen above the critical kicking strength $\lambda_c^{\rm N(S)} \sim 0.393 (0.556)$.}
\label{q_scar}
\end{figure}

To this end, we investigate the presence of quantum scars which can appear due to the fixed points of the Hamiltonian map, and analyze their dynamical signature from the OTOC. As a natural choice for demonstration, we consider two trivial fixed points of KDM namely, the north pole $(s_+^*)$ and south pole $(s_-^*)$ (cf. Sec.\ \ref{recap}), which remain stable until the kicking strength $\lambda_0$ is increased to their respective critical values, $\lambda^{\rm N}_c$ and $\lambda^{\rm S}_c$ respectively. It is a pertinent question to ask, whether the instability of these fixed points lead to an ergodic dynamics, or  does the trace of these fixed points give rise to a deviation from ergodic behavior in the local phase space region? To answer this question, we study the dynamics of OTOC in the spin sector, $C_{S_z}(n)$, using the coherent states $\vert s_{\pm}^* \rangle$ corresponding to the unstable fixed points $s_{\pm}^*$ in the phase space, 
and contrasted it with a typical point within a local chaotic region of the mixed phase space at a given kicking strength $\lambda_0$, as shown in Fig.\ \ref{q_scar}(a,b).
For the purpose of comparison, we scale $C_{S_z}(n)$ by the saturation value of the correlator $\bar{C}^{\rm GOE}_{S_z}$, obtained from the time evolution under $\hat{\mathcal{U}}_{\rm GOE}$.
Clearly, the growth and magnitude of the scaled correlator at the unstable fixed points $s_{\pm}^*$, remain significantly small as compared to a typical chaotic point. It is important to mention that, such a significant effect of scarring can only be observed in the mixed phase region for intermediate kicking strengths. Whereas, $C_{S_z}/\bar{C}^{\rm GOE}_{S_z} \rightarrow 1$ in the completely ergodic regime, indicating the equivalence of all the phase space points, where the effect of scarring can be lost.

To further elucidate the effect of scarring due to the unstable fixed points, we compute the overlap, $|\langle\Phi_{\nu}|s^{*}_{\pm}\rangle|^2$ between the corresponding coherent state $\vert s_{\pm}^* \rangle$ and the Floquet eigenstates $\vert \Phi_{\nu} \rangle$. It is evident from Fig.\ \ref{q_scar}(c,d), there are few Floquet eigenstates which have a significant overlap with the coherent states $ $ even for $\lambda_0>\lambda^{\rm N,S}_c$. Whereas, this behavior can be contrasted with a typical point from chaotic region of the phase space, for which the corresponding coherent state is delocalized over the Floquet eigenbasis. We consider those Floquet eigenstates, which have a maximum overlap with $\vert s^{*}_{\pm} \rangle$ and analyze their semiclassical phase space distribution by computing the Husimi distribution,
\begin{equation}
Q(\theta,\phi) = \frac{1}{\pi}\langle\theta,\phi|\hat{\rho}^{\nu}_S|\theta,\phi\rangle
\end{equation}
Such phase space distribution over the Bloch sphere clearly shows a significant accumulation of density near the north and south poles [see Fig.\ \ref{q_scar}(e,f)] respectively, which can be identified as the scar of the unstable fixed points. In this context, we would like to mention that the scarring of the north pole ($s^{*}_+$), is similar to the scarring of $\pi$-mode in the Bose-Josephson junction \cite{Sinha20}. As revealed from the analysis, the effect of scarring leads to a localization in phase space region, which hinders the ergodic behavior of the OTOC dynamics.

\section{Summary}
\label{conclu}

In summary, we have studied the onset of chaos and quantum scars in a periodically kicked Dicke model, using the growth and long time saturation of out-of-time-order correlators (OTOC) as a diagnostic tool. By constructing the Hamiltonian map in the classical limit achieved for $S \rightarrow \infty$, we have analyzed the fixed points and their stability, as well studied the classical phase space dynamics which changes from a regular to chaotic behavior by tuning the kicking strength $\lambda_0$. The onset of chaos is also reflected in the spectral statistics of the Floquet operator. With increasing $\lambda_0$, the distribution of quasi-energy spacings changes from Poisson to Wigner-Dyson statistics belonging to the orthogonal class of RMT.
As an alternate measure to quantify the degree of chaos, we studied the OTOC dynamics for the oscillator as well in the spin sector of KDM. In order to probe chaos locally in the phase space, we investigated the behavior of the OTOCs by preparing the system initially in the coherent state representing a phase space point. The growth rate of the OTOC constructed from the canonically conjugate variables of the oscillator, agrees with the Lyapunov exponent reasonably well in the chaotic regime. Whereas, in the mixed phase space, the deviation between these two quantities becomes significant. The OTOC constructed in the spin sector also exhibits a similar growth in the initial time scale followed by a saturation beyond the Ehrenfest time. In the regular regime, the magnitude of such a saturation value remains vanishingly small. Interestingly, as $\lambda_0$ is increased, the saturation value of OTOC in the spin sector evaluated with a microcanonical ensemble increases, and finally attains its maximum value in the deep chaotic regime. Such a maximum can also be obtained by considering the time evolution under a random unitary matrix of orthogonal class. 
Furthermore, the behavior of the saturation value of the OTOC in the crossover regime has been captured from a random matrix model by mixing a Poisson and a random matrix belonging to orthogonal ensemble. 
The present analysis reveals that the saturation value of the OTOC can serve as an alternate measure to detect the crossover to chaos in an interacting quantum system. Such a measure has also been used for analyzing delocalization transition of bosons in a driven quasiperiodic potential \cite{Ray18}.

We further identify the quantum scars of fixed points of the KDM, particularly the trivial fixed points, corresponding to the north and south pole of the Bloch sphere. Even in the region, where the above mentioned fixed points become unstable, their signature is retained in the Floquet eigenstates, which can be seen from the Husimi-distribution. The suppression in growth of OTOC dynamics, evaluated for the coherent states corresponding to those unstable fixed points is a clear evidence of deviation from ergodicity due to the effect of scarring. Such states have appreciable overlap with a few Floquet eigenstates containing the scars, whereas the overlap of a coherent state corresponding to an arbitrary point in the chaotic region of phase space is delocalized over the Floquet basis states. Such localization behavior due to presence of quantum scars hinder the ergodic evolution of the correlators. 

In conclusion, we have analyzed both classically and quantum mechanically, the onset of chaos in kicked Dicke model as an example to demonstrate that, the saturation value of OTOC apart from its growth rate can serve as an alternate measure to quantify the degree of chaos in an interacting quantum system. Moreover, the OTOC dynamics can also be used to identify the quantum scars. The model and the results of the present work have experimental relevance, since Dicke model has already been realized by coupling a condensate with the cavity mode \cite{Esslinger10, Esslinger13, Hemmerich15}. 
Such a model can also be realized in Bose-Josephson junction \cite{Gati06} and circuit QED setup \cite{Paraoanu19}. Finally, measurement of OTOC in spin sector can be performed following the similar line of thought, which has been implemented previously for NMR system and in trapped ions \cite{Rey17, Du17}.

\section*{Acknowledgement}
SR acknowledges the scholarship of the Alexander von Humboldt Foundation, Germany.

\appendix

\section{Effective random matrix model}
\label{Eff_RMT}

In this appendix, we elaborate a system independent study to understand the change in the saturation value of the OTOC of spin variable as the crossover from regular to ergodic regime occurs. Generically, such a crossover between the two dynamical regimes is associated with a change in the underlying spectral statistics, which can be captured from an effective random matrix model \cite{Brody_rmp,Wettig12}, constructed from a mixture of two random matrices belonging to the Poisson and GOE class. The model Hamiltonian can be described by the following matrix, 
\begin{equation}
\hat{H}_{\rm RMT} = \hat{H}_{\rm P} + \frac{\alpha}{\mathcal{\sqrt{D}}} \hat{H}_{\rm GOE}
\end{equation}
where $\alpha$ is a mixing parameter which controls the crossover from Poisson ($\hat{H}_{\rm P}$) to GOE ($\hat{H}_{\rm GOE}$) random matrix. We consider the random matrix $\hat{H}_{\rm RMT}$ of dimension $\mathcal{D}=(2S+1)$ to calculate the spin OTOC $\bar{C}_{S_z}$. The resulting Floquet operator is given by $\hat{\mathcal{U}}_{\rm RMT}= e^{i\hat{H}_{\rm RMT}T}$, and its eigenspectrum is characterized by the eigenphases $\phi_{\nu}$ and the corresponding eigenvectors $\vert \Phi_{\nu} \rangle$. 

\begin{figure}[t]
\centering
\includegraphics[clip=true,width=\columnwidth]{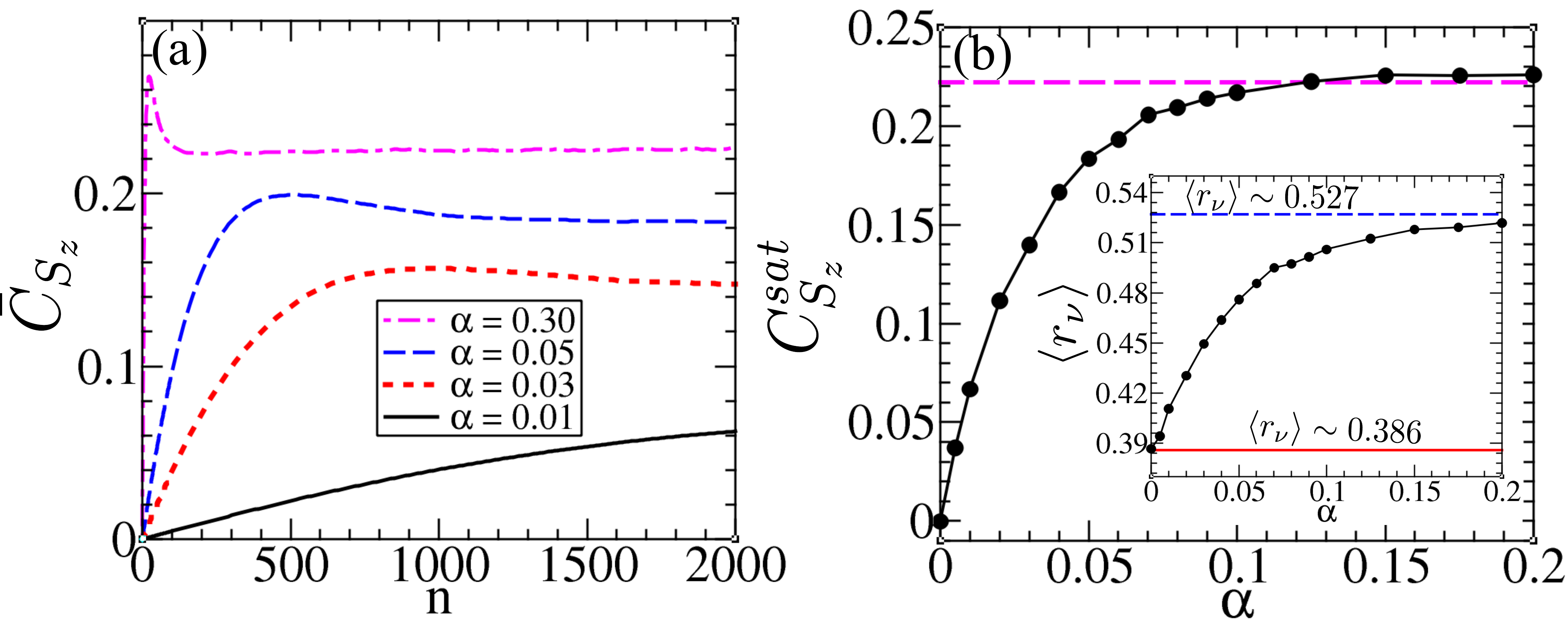}
\caption{{\it Effective RMT model:} (a) Stroboscopic dynamics of OTOC $\bar{C}_{S_z}(n)$ for different mixing parameter $\alpha$. (b) Saturation value of spin OTOC, $C_{S_z}^{sat}$, with increasing $\alpha$. The pink dashed line represents the saturation value obtained by using $\hat{H}_{\rm GOE}$. The inset shows the average level spacing ratio as a function of $\alpha$. Parameters chosen: $S = 500$ and $T=1$.}
\label{eff_rmt}
\end{figure}

In Fig.\ \ref{eff_rmt}, the behavior of OTOC is portrayed as the spectral statistics changes from Poisson to GOE, mimicking the crossover from regular to chaotic dynamics in the KDM, where the tuning parameter $\alpha$ of RMT plays a similar role as the kicking strength $\lambda_0$. First, we compute $\bar{C}_{S_z}(n)$ in the microcanonical ensemble using the eigenvectors of $\hat{\mathcal{U}}_{\rm RMT}$ [see Eq.\ \eqref{OTOC_sx}], as shown in Fig.\ \ref{eff_rmt}a. In the long time, $\bar{C}_{S_z}(n)$ saturates, and the saturation value $C_{S_z}^{sat}$ increases with increasing $\alpha$, which finally attains a maximum limit for larger $\alpha$, as depicted in Fig.\ \ref{eff_rmt}b. Such an upper limit can be obtained by using the random matrix of GOE class i.e. $\hat{H}_{\rm RMT} = \hat{H}_{\rm GOE}$, which we have marked by a horizontal dashed line in Fig.\ \ref{eff_rmt}b. For completeness, we also plot the average level spacing ratio $\langle r_{\nu} \rangle$ [see Eq.\ \eqref{r_avg}] obtained from $\phi_{\nu}$, that exhibits a crossover from Poisson to orthogonal class of RMT, as $\alpha$ is increased (see the inset of Fig.\ \ref{eff_rmt}b). Therefore, an increase in the saturation value of OTOC, associated with the onset of chaotic dynamics in KDM, is more generically captured from the random matrix model, as discussed above. Thus, such a behavior of the saturation value of OTOC can serve as a tool to identify the onset of underlying chaos in a generic interacting quantum system.  

\end{document}